\documentclass{aa}
\usepackage{graphics}

\newcommand{\ms}{m\,s$^{-1}$} 
\newcommand{\kms}{km\,s$^{-1}$} 
\newcommand{\bz}{$\langle B_{\rm z} \rangle$}  
\newcommand{\equ}{$\gamma$\,Equ}

\newcommand{\pr}{\ion{Pr}{iii}}
\newcommand{\nd}{\ion{Nd}{iii}}

\newcommand{\dfrac}[2]{\frac{\displaystyle #1}{\displaystyle #2}}

\newcommand{\fifps}[2]{\centering\resizebox{#1}{!}{\includegraphics{#2}}}

\begin{document}

\title{No magnetic field variation with pulsation phase \\ in the roAp star $\gamma$~Equulei}

\author{O. Kochukhov\inst{1} \and T. Ryabchikova\inst{2,1} \and N. Piskunov\inst{3}}

\offprints{O. Kochukhov \\ \email{kochukhov@astro.univie.ac.at}}

\institute{Institut f\"ur Astronomie, Universit\"at Wien, T\"urkenschanzstra{\ss}e 17, 1180 Wien, Austria
      \and Institute of Astronomy, Russian Academy of Sciences, Pyatnitskaya 48, 109017 Moscow, Russia
      \and Uppsala Astronomical Observatory, Box 515, SE-751 20 Uppsala, Sweden}

\date{Received / Accepted }

\abstract{We present an analysis of 210 high-resolution time-resolved spectropolarimetric
observations of the roAp star \equ\ obtained over three nights in August and September 2003. Radial
velocity variations due to {\it p-}mode non-radial pulsations are clearly detected in the
lines of rare-earth elements, in particular \pr, \ion{Nd}{ii} and \nd. In contrast, we find 
absolutely no evidence for the variation of the mean longitudinal magnetic field over the 
pulsation period in \equ\ at the level of 110--240~G which was
recently reported by Leone \& Kurtz (\cite{LK03}). Our investigation of the variability of
circularly polarized profiles of 13 \nd\ lines demonstrates that, at the 3$\sigma$
confidence level, no magnetic field variation with an amplitude above $\approx40$--60~G was
present in \equ\ during our monitoring of this star.
\keywords{stars: atmospheres -- stars: chemically peculiar -- stars: individual: \equ\ --
stars: magnetic fields -- stars: oscillations}}

\maketitle

\section{Introduction}
\label{intro}

Since the discovery of high-overtone {\it p-}mode pulsations in the rapidly oscillating Ap (roAp) stars
it became clear that, unlike in other stellar pulsators, strong magnetic fields in  these stars have a
defining role in exciting the oscillations and shaping main pulsational properties. It was found that
the amplitude and phase of the rapid light variation are modulated by the stellar rotation and that the
phases of the extrema of magnetic field and pulsational amplitude typically coincide with each other. A
phenomenological oblique pulsator model (Kurtz \cite{K82}) attributed main characteristics of the
photometric pulsational variation of roAp stars to the oblique dipole modes, aligned with the axis of
dipolar magnetic field. 

The apparent connection between global magnetic field of roAp stars and their non-radial pulsations
led Hubrig et al. (\cite{HKB03}) to propose that atmospheric magnetic structure may also change over
the pulsation period and hence pulsational variability of the longitudinal magnetic field can
eventually be observed. Hubrig et al. (\cite{HKB03}) searched for the magnetic variations in a sample of
six roAp stars but failed to detect any coherent magnetic variability in low-resolution Zeeman
spectra. This null result is not unexpected given that in roAp stars lines of different elements
display dramatically different amplitudes and phases of oscillations, which are combined
destructively and result in an extremely weak pulsational signal when averaged over broad wavelength
regions. 

High-resolution analysis of Stokes profiles of individual metal lines is a much more promising
tool for mapping atmospheric pulsational disturbances and investigating possible oscillations of
magnetic field. In a recent paper Leone \& Kurtz (\cite{LK03}, hereafter LK) announced the
discovery of variation of the mean longitudinal field \bz\ with the 12.1-minute pulsation period
in one of the brightest roAp star \equ. Magnetic variability with an amplitude between 110 and
240~G was inferred from the study of 4 \nd\ lines in only 18 circularly polarized spectra of \equ,
which makes the alleged detection of magnetic variation a statistically marginal result. With the
aim of verifying the existence of rapid magnetic variability in \equ\ we decided to acquire and
analyse a much more extensive time series spectropolarimetric  observations of this star.

\section{Observations}
\label{obser}

We obtained time-resolved circularly polarized spectra of \equ\ using the cross-dispersed Nasmyth
Echelle Spectrometer (NES, Panchuk et al. \cite{NES}) installed at the 6-m  telescope of the Russian
Special Astrophysical Observatory. The NES instrument is equipped with a Zeeman analyzer consisting of
a quarter-wave and half-wave retarders and a calcite plate separating the beams with opposite circular
polarization. A 2K$\times$2K Loral CCD detector allowed us to record 26 echelle orders with nearly
complete wavelength coverage in the 4520--6000~\AA\ region with a spectral resolving power of
$R\approx38\,000$.

On each of the nights of August 19 and 20, 2003 we observed \equ\ for 2.4$^{\rm h}$ and have  obtained
continuous sequences of 70  Stokes $I$ and $V$ spectra using 80$^{\rm s}$ integration time with a
readout time of 42$^{\rm s}$. An additional series of 70 polarized spectra of \equ\ was secured on
September 11, 2003, when exposure times of 120$^{\rm s}$ were used. On each of the observing nights the
peak $S/N\approx60$--80 was reached in individual time-resolved spectra. The software package {\sl
REDUCE} (Piskunov \& Valenti \cite{PV02}) was used for the optimal extraction of the NES spectra. A
wavelength scale with an internal accuracy of 50--70~\ms\ was established by means of a 2-D wavelength
calibration procedure, which made use of $\approx$700 ThAr lines in all echelle orders. The left- and
right-hand circularly polarized (LCP and RCP) beams were calibrated independently using the
corresponding beams of the reference spectrum taken before and after stellar observations.

\section{Identification of \nd\ lines}
\label{nd3}

Investigation of pulsational variability in the spectral lines of the rare-earth elements (REE) offers
unique potential to observe non-radial {\it p-}mode oscillations and to study in detail the structure of
pulsating cavity in roAp stars. It was recently demonstrated by Kochukhov \& Ryabchikova
(\cite{KR01a,KR01b}) that the largest pulsational variability (RV amplitudes up to 1~\kms) in
individual metal lines in \equ\ and other roAp stars is invariably found in the REE spectral lines, in
particular strong absorption features of doubly ionized Pr and Nd, while pulsations could not be
detected in the lines of light and iron-peak elements. Since the lines of each REE ion show very
similar pulsational behaviour, the accuracy of our search for magnetic oscillations in \equ\ can be
substantially improved by simultaneous multiline analysis of \nd\ lines, which constitute the most
numerous group of strong lines with pronounced variability in the spectrum region available in our
observations.

In the present study magnetic field and RV variations were measured using 13 unblended \nd\ lines some
of which were previously unclassified, but appeared in laboratory spectra (Crosswhite \cite{C76};
Ald\'enius \cite{MA01}) as well as in the spectra of roAp stars known to have \nd\ lines of abnormal
intensity (Ryabchikova et al. \cite{RSMK01}). We emphasize that all newly classified \nd\ lines show
strong pulsational signatures, consistent with the behaviour of known lines of \nd, which by itself
provides a convincing support for our line identification. Table~\ref{tbl1} contains wavelengths of
\nd\ lines, proposed transition identification based on the energy level calculations by Zhang et al.
(\cite{ZSP02}) and the effective Land\'e factors which were calculated from the $g$-factors published
by Bord (\cite{DB00}) and Zhang et al. (\cite{ZSP02}). The details of the line classification will be
presented elsewhere, here we note that the adopted Land\'e factors and Zeeman patterns are consistent
with the observed Zeeman splitting in the sharp-lined Ap star HD\,144897 with the surface magnetic
field of $\approx$ 9 kG.

\begin{table} 
\caption{\nd\ lines used in the RV and magnetic field analysis of \equ. The columns
give laboratory wavelength, effective Land\'e factor, excitation
potentials and $J$ values of the lower and upper levels. \label{tbl1}}
\begin{center}
\begin{tabular}{lcc}
\noalign{\smallskip}
\hline
\hline
~~~$\lambda$ (\AA)          &$g_{\rm eff}$ & Transition (cm$\,^{-1}$) \\
\hline
4759.54$^{bc}$& 1.621   & 5093$_8$ -- 26098$_7$\\
4796.49$^{bc}$& 1.553   & 1138$_5$ -- 21981$_6$\\
4927.49$^a$   & 1.186   & 3714$_7$ -- 24003$_7$\\
4942.68$^b$   & 1.479   & 3714$_7$ -- 23941$_6$\\
5012.94$^d$   & 1.677   & 5093$_8$ -- 25036$_7$\\
5203.92$^{ad}$& 0.919   & 1138$_5$ -- 20349$_5$\\
5286.75$^a$   & 1.421	& 5093$_8$ -- 24003$_7$\\
5294.11$^a$   & 0.615	&~~~~~~0$_4$ -- 18884$_4$\\
5677.18$^a$   & 1.579	& 5093$_8$ -- 22703$_7$\\
5802.54$^c$   & 1.531	& 2387$_6$ -- 19617$_5$\\
5845.02$^a$   & 1.101	& 5093$_8$ -- 22197$_9$\\
5851.54$^c$   & 1.617	& 3714$_7$ -- 20799$_6$\\
5987.68$^a$   & 1.163	& 3714$_7$ -- 20411$_7$\\
\hline
\end{tabular}
\end{center}
$^a${\footnotesize{Previously identified and classified \nd\ lines}}\\
$^b$Lines included in Crosswhite's (\cite{C76}) unpublished list\\
$^c$Lines detected in the laboratory spectrum (Ald\'enius \cite{MA01} and private communication) \\
$^d$Reclassified and newly classified lines
\end{table}

\section{Radial velocity and magnetic field measurements}
\label{rvmag}

\begin{figure*}[!th]
\fifps{9cm}{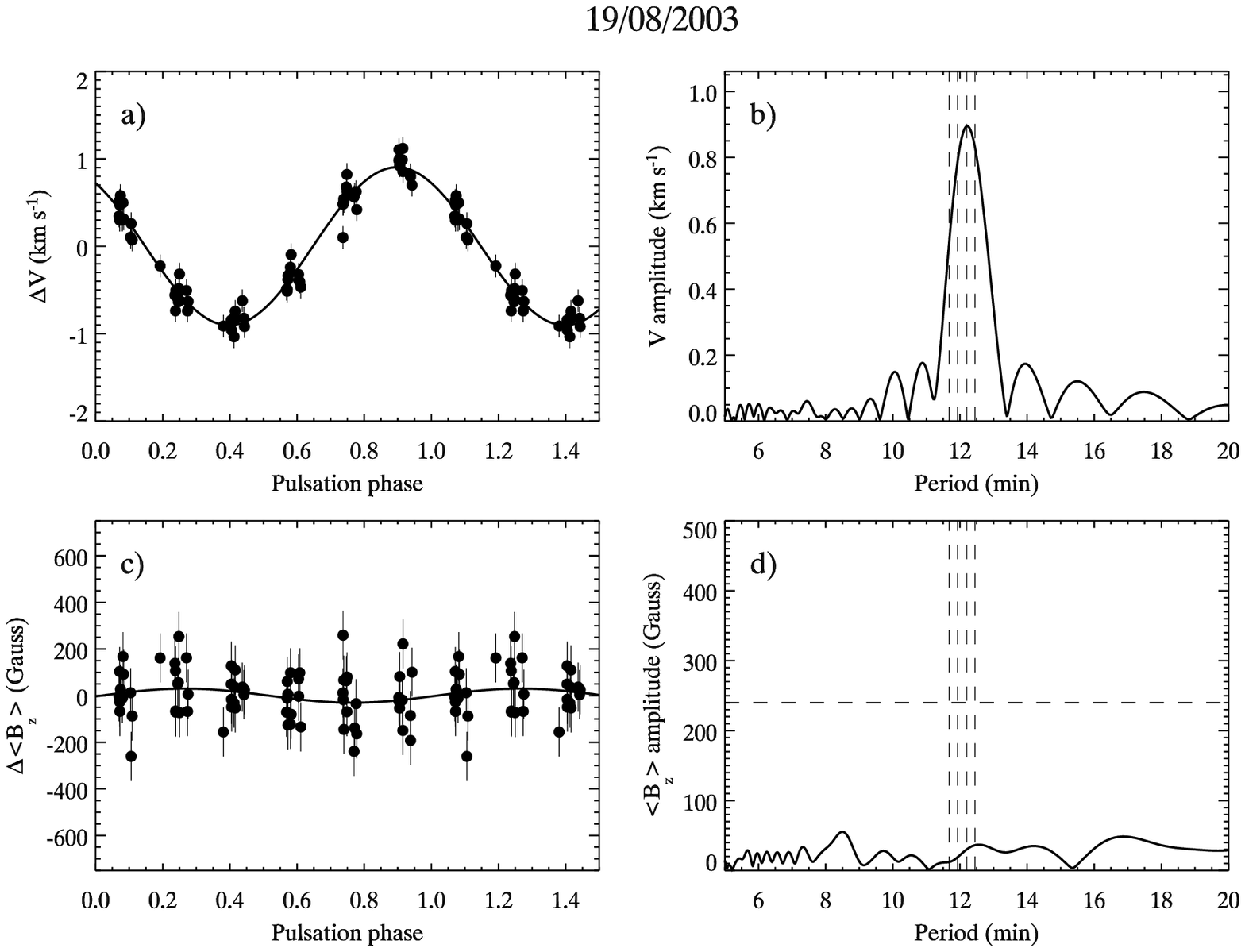}\fifps{9cm}{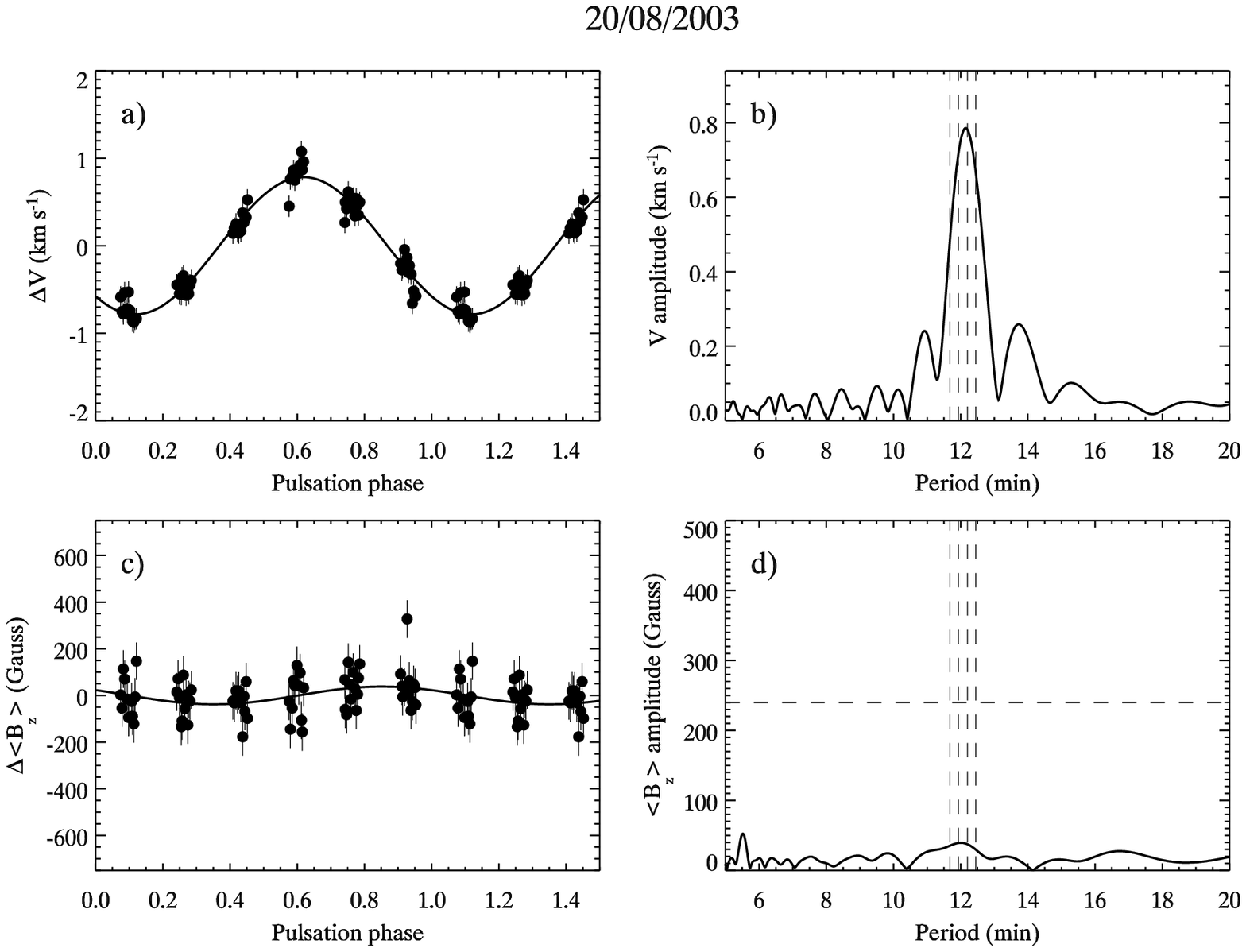}
\caption{Pulsational variation of the radial velocity and longitudinal magnetic field measured
for \nd\ lines on the nights of August 19 and 20, 2003. The average pulsational RV curves
{\bf a)} are folded with the best-fit oscillation period and are compared with the
average variation of \bz\ {\bf c)}. Panels {\bf b)} and {\bf d)} illustrate the amplitude
spectra for the RV and longitudinal magnetic field. The vertical dashed lines show
photometric  pulsational periods of \equ. The horizontal dashed line in panels {\bf d)}
indicates the maximum amplitude  of pulsational \bz\ variation reported by LK.}
\label{fig1}
\end{figure*}

\begin{table*}[th]
\centering
\caption{Results of our analysis of the time-resolved polarized spectra of \equ. The columns list UT
date of observation, reference Julian date, number of analysed spectra and a least-squares estimate of
the pulsational period, average amplitude and phase (measured in fractions of the period) of the pulsational RV variation
in 13 \nd\ lines. The last  three columns report the amplitude of the magnetic variation, standard
deviation of the \bz\ measurements and an amplitude which would have been detected with out data at the
3$\sigma$ confidence level.  \label{tbl2}}
\begin{tabular}{clclccccc}
\hline
\hline
UT Date & $T_0-$   & N & ~~~~~~$P$  & $\langle A\rangle$ & $\varphi$ & $\Delta$\bz\ & $\sigma($\bz$)$& $\Delta$\bz$_{\rm max}$\\ 
        & 2400000  &   & ~~~(min)& (\ms)                 &           & (G)          & (G)            & (G)             \\
\hline
\multicolumn{9}{c}{13 lines of \nd} \\
19/08/2003 & 52871.5140 & 70 & 12.212$\pm$0.013 & 963       &  0.102$\pm$0.004 & 30$\pm$17 & 105 & 63 \\
20/08/2003 & 52872.3727 & 70 & 12.169$\pm$0.012 & 854       &  0.382$\pm$0.004 & 37$\pm$13 &  80 & 48 \\
11/09/2003 & 52894.2954 & 70 & 12.231$\pm$0.024 & 262       &  0.011$\pm$0.012 & 14$\pm$10 &  59 & 36 \\
\hline
\end{tabular}
\end{table*}

The pulsational analysis of \equ\ followed the procedure outlined in Kochukhov \& Ryabchikova
(\cite{KR01a}). For each spectral line in the available wavelength region we determined the
center-of-gravity as a function of time in the Stokes $I$ spectra. Then a refined analysis of
polarization spectra was carried out for the group of \nd\ lines aimed at detection of the magnetic
variability. An instantaneous position of the line center was determined consistently for the RCP and
LCP profiles by integrating over the spectral points below certain residual depth. The same vertical
cutoff was applied to the RCP and LCP spectra, but it had to be chosen individually for each line due
to different blending in the far wings. This results in up to a factor of two difference in the RV
amplitudes for studied \nd\ lines, although depth effects may also be partially responsible for that.
The RV and \bz\ were derived from the center-of-gravities of the RCP and LCP profiles, $\lambda_{\rm
R}$ and $\lambda_{\rm L}$, using the expressions
\begin{equation}
\begin{array}{rl}
\langle B_{\rm z} \rangle = & \dfrac{\lambda_{\rm R} + \lambda_{\rm L}}
                       {2.34 \times 10^{-13} g_{\rm eff} (\langle\lambda_{\rm R}\rangle + \langle\lambda_{\rm L}\rangle)^2}, \\
                       RV = & c \left(\dfrac{\lambda_{\rm R} + \lambda_{\rm L}}{
		       \langle\lambda_{\rm R}\rangle + \langle\lambda_{\rm L}\rangle} - 1\right),
\end{array}
\end{equation} 
where $c$ is the speed of light and $\langle\lambda_{\rm R,L}\rangle$ correspond to the time-averaged line center
positions.

\begin{figure}[!th]
\fifps{\hsize}{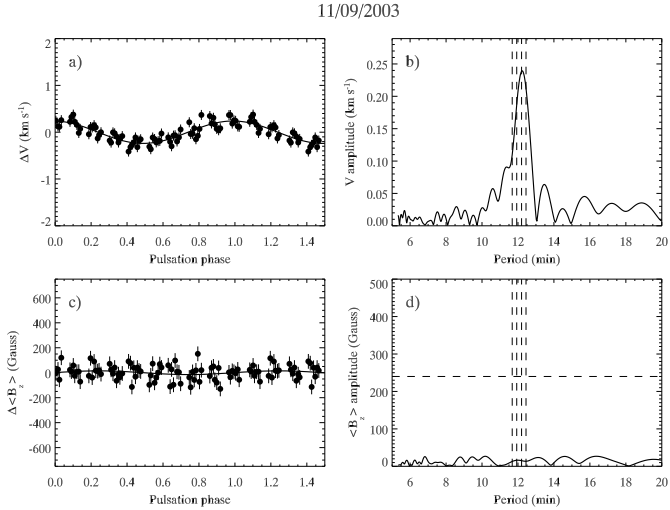}
\caption{The same as Fig.~\ref{fig1} for the spectropolarimetric observations of \equ\
obtained on the night of September 11, 2003.}
\label{fig2}
\end{figure}

Preliminary time series analysis of the \nd\ lines showed negligible difference between the periods
and phases of their RV variability. Therefore, in subsequent least squares sinusoidal fits we
allowed for individual amplitudes, but used the same phase and pulsation period for all \nd\
lines, effectively coadding information from many lines into a single high $S/N$ RV measure. The
same strategy was used to construct the average variation of the longitudinal field after
subtracting a constant offset from \bz\ measurements obtained for each of the \nd\ lines. 
Results of our investigation of the neodimium line variability are summarized in Table~\ref{tbl1}
and are illustrated in Figs.~\ref{fig1} and \ref{fig2}, which show the average RV curves (scaled to
nightly mean amplitudes) as well as the average variation of \bz\ and corresponding amplitude
spectra.

For all three nights the pulsation period detected in our data is consistent with one of the
frequencies ($P=12.201$~min) identified in photometry (Martinez et al. \cite{MART96}). Despite
unambiguous detection of the pulsational RV shifts, no evidence for magnetic variability is found.
A rigorous statistical analysis indicates that, at the 3$\sigma$ confidence level, no magnetic
variability with $\Delta\langle B_{\rm z}\rangle > 40$--60~G is seen in \equ\ during the nights
of our observations.

In addition to the study of the \nd\ lines, we identified a number of \pr\ and \ion{Nd}{ii}
features showing clear pulsational variability. These lines are not so numerous and strong as to
provide an independent and accurate diagnostic of possible rapid magnetic changes. Nevertheless,
analysis of these ions is consistent with the results obtained using the \nd\ lines: two lines of
\pr\ with the average RV amplitude between 950~\ms\ (19 Aug) and 280~\ms\ (11 Sep) show 
$\Delta\langle B_{\rm z}\rangle \la 100$~G, while  $\Delta\langle B_{\rm z}\rangle \approx
60\pm50$~G is derived from four \ion{Nd}{ii} lines. Remarkably, the lines of singly ionized Nd,
which could not be investigated in our first spectroscopic study of pulsations in \equ\ (Kochukhov
\& Ryabchikova \cite{KR01a}), show very high average RV amplitudes of 1416~\ms\ (19 Aug),
1118~\ms\ (20 Aug) and 414~\ms\ (11 Sep) and a constant phase shift of $0.13\pm0.02$ of pulsation
period with respect to the RV variation of \nd.

The procedure of the magnetic and RV measurements applied to REE lines was also tested on a sample of 19
unblended strong lines of Ca, Cr and Fe, which are not expected to show any pulsational variability of
either RV or \bz. From the Aug 19 data we obtained RV amplitude of $30\pm19$~\ms, 
$\Delta\langle B_{\rm z}\rangle = 18\pm14$~G and standard deviation $\sigma($\bz$)=84$~G for these lines, in good quantitative
agreement with the REE results. This provides a complementory indirect argument for the absence of
magnetic variability in REE lines. Since the variation of magnetic field is predicted to be proportional
to the RV oscillations (see Hubrig et al. \cite{HKB03}), even for marginal magnetic variation one would
expect to find $\sigma($\bz$_{\rm REE}) \gg  \sigma($\bz$_{\rm Ca,Cr,Fe})$, but this is clearly not
observed in \equ. 

\section{Discussion}
\label{disc}

What may be the reason for a major discrepancy between the results reported by LK and the outcome
of our study of \equ? We consider it improbable that the star itself has changed its
pulsational behaviour dramatically. Theoretical estimate of possible magnetic variability (Hubrig
et al. \cite{HKB03}) hints that magnetic and velocity amplitudes are proportional to each other.
Therefore, if magnetic variability really exists, on the nights of Aug 19 and 20 (when RV 
amplitudes of \nd\ lines were about twice the amplitude measured by LK) we should have observed
even larger \bz\ variation than found by LK.

\begin{figure}[!th]
\fifps{4.3cm}{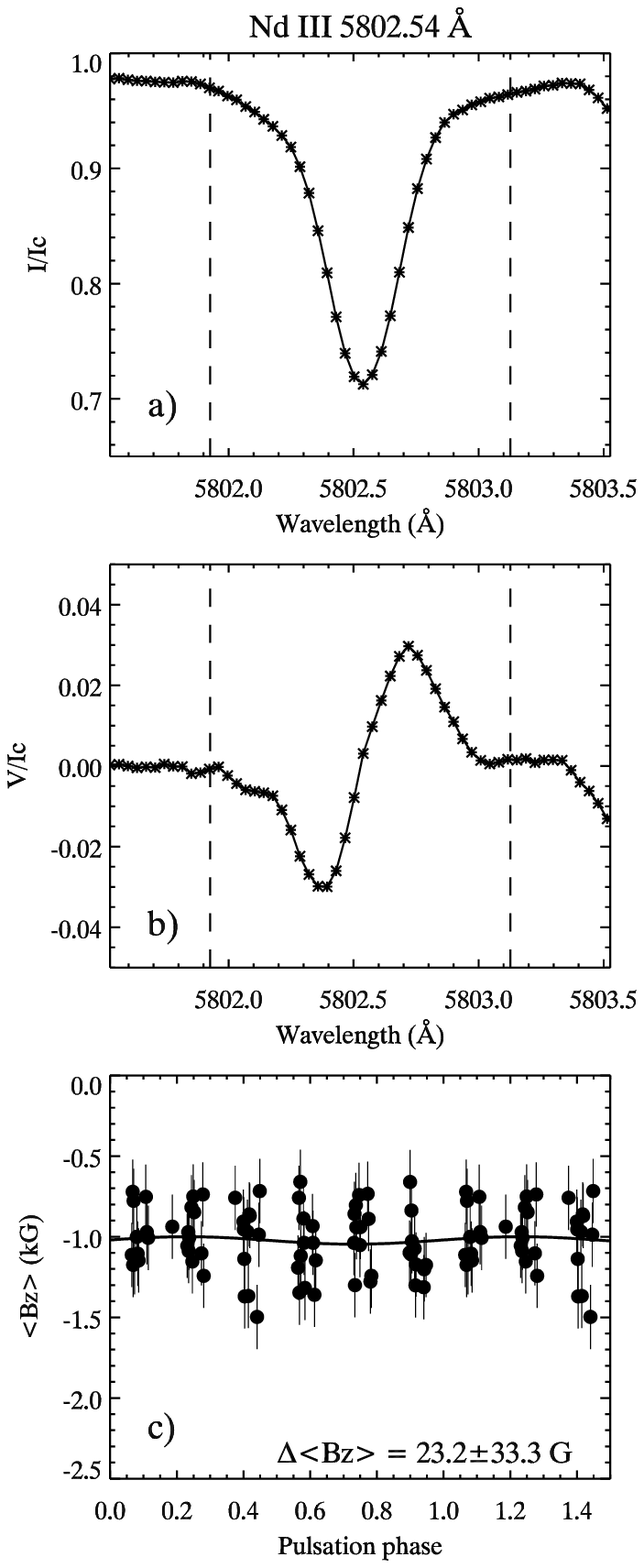}\hspace*{0.1cm}\fifps{4.3cm}{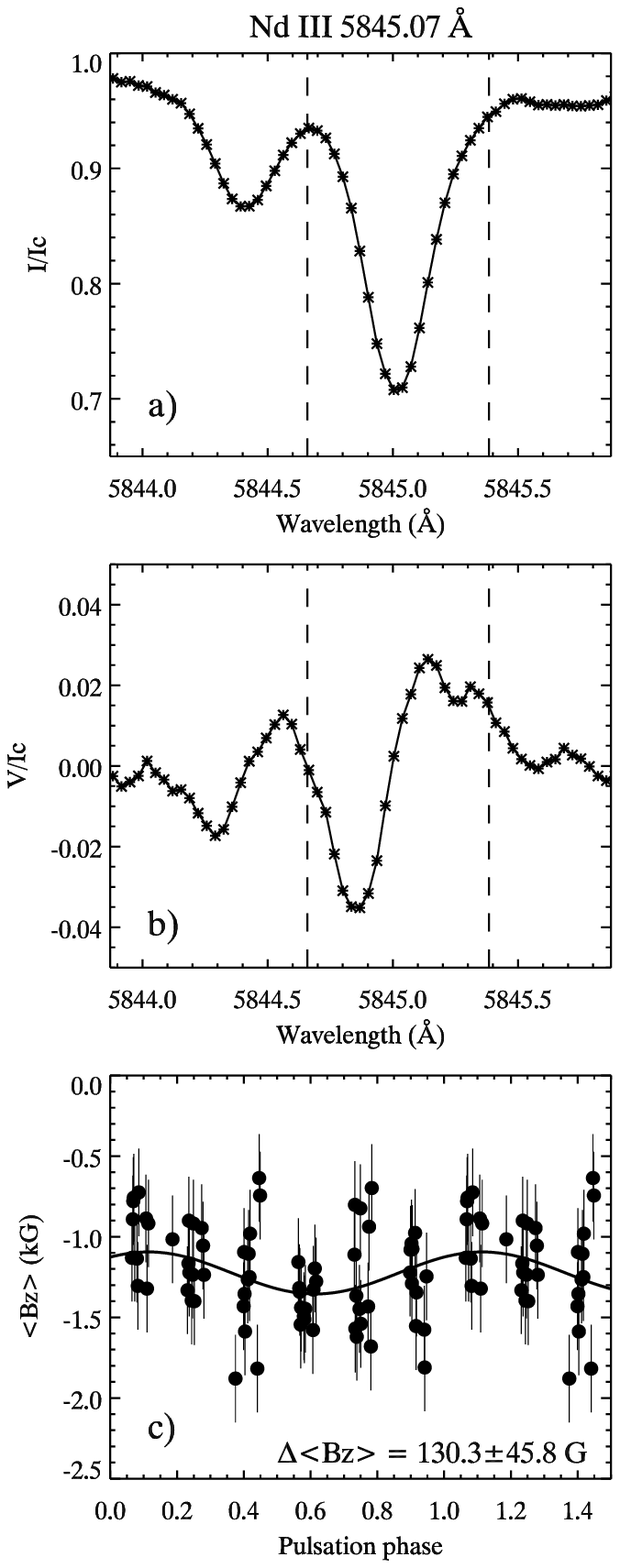}
\caption{The average Stokes $I$ {\bf a)} and $V$ {\bf b)} profiles of the \nd\ lines at 
$\lambda$ 5802.54 and 5845.07~\AA. The vertical dashed lines illustrate selection of 
the spectral region for the RV and magnetic measurements. The panel {\bf c)} shows variation of the
\bz\ measure obtained from the Stokes $V$ profiles. Blending by the \pr\ 5844.41
\AA\ leads to spurious variation of \bz\ determined for the \nd\ 5845.07~\AA.}
\label{fig3}
\end{figure}

It is possible that the difference between our null result and the alleged positive detection of the
magnetic variability by LK is related to different approaches to measure \bz\ and the effects of
blending of the \nd\ lines studied by LK. We derive longitudinal field from consistent
center-of-gravity determination in the RCP and LCP profiles, while LK analysed the first moment of
the Stokes $V$ spectra. Strong \nd\ lines tend to show extended wings and have unusually wide
Stokes $V$ signatures. Hence, an insignificant blending in Stokes $I$ may correspond to a
substantial overlap of the Stokes $V$ profiles of neighbouring lines. The \nd\ 5845.07~\AA, which
LK offered as the most prominent case of \bz\ variation, represents a clear example of the
blending problem. This line is blended by the \textit{variable} line of \pr\ at $\lambda$
5844.41~\AA, which appears to be reasonably well separated from the \nd\ line in Stokes $I$, but
significantly distorts the corresponding Stokes $V$ signature. In Fig.~\ref{fig3} we use our Aug
19 data to illustrate that the blending of the \nd\ 5845.07~\AA\ results in spurious
variation of \bz\ when integrating Stokes $V$ signal for this line within a fixed wavelength range. In contrast,
no variation is seen in the nearby line of \nd\ 5802.54~\AA\ which is free from any blends and has
higher magnetic sensitivity.

We also point out that the results of the RV and magnetic measurements presented by LK for
individual \nd\ lines are inconsistent. The reported phase shift between \bz\
curves of, e.g., the \nd\ 5845.07 and 6145.07~\AA, reaches 0.20 of the pulsation period (with a
typically phase error not exceeding 0.03), while no such shift is seen in RV variation.
Unfortunately, LK do not comment on this peculiar behaviour of different \nd\ lines.  

Based on our observations we conclude that we see no evidence for pulsational variation of 
magnetic field in \equ\ contrary to the recent claim by LK. We show that systematic
problems exist in the latter study and hence the alleged detection of the magnetic variability in
\equ\ may be spurious.

\begin{acknowledgements}
We sincerely thank V.~Panchuk and M.~Yushkin for carrying out the observations of \equ.
This work was supported by the Lise Meitner fellowship to OK (FWF  project M757-N02), FWF project $P
14984$ and has been partly funded by the Swedish National Research Council, Swedish Royal Academy of
Sciences and by the Russian Federal program ``Astronomy'' (part 1102).
\end{acknowledgements}

\end{document}